\PassOptionsToPackage{unicode}{hyperref}
\PassOptionsToPackage{hyphens}{url}
\documentclass[
  11pt,
]{article}
\usepackage{xcolor}
\usepackage[margin=1in]{geometry}
\usepackage{amsmath,amssymb}
\usepackage{iftex}
\ifPDFTeX
  \usepackage[T1]{fontenc}
  \usepackage[utf8]{inputenc}
  \usepackage{textcomp} 
\else 
  \usepackage{unicode-math} 
  \defaultfontfeatures{Scale=MatchLowercase}
  \defaultfontfeatures[\rmfamily]{Ligatures=TeX,Scale=1}
\fi
\usepackage{lmodern}
\ifPDFTeX\else
\fi
\IfFileExists{upquote.sty}{\usepackage{upquote}}{}
\IfFileExists{microtype.sty}{
  \usepackage[]{microtype}
  \UseMicrotypeSet[protrusion]{basicmath} 
}{}
\makeatletter
\@ifundefined{KOMAClassName}{
  \IfFileExists{parskip.sty}{%
    \usepackage{parskip}
  }{
    \setlength{\parindent}{0pt}
    \setlength{\parskip}{6pt plus 2pt minus 1pt}}
}{
  \KOMAoptions{parskip=half}}
\makeatother
\usepackage{color}
\usepackage{fancyvrb}

\DefineVerbatimEnvironment{Highlighting}{Verbatim}{commandchars=\\\{\}}
\newenvironment{Shaded}{}{}

\newcommand{\AttributeTok}[1]{\textcolor[rgb]{0.49,0.56,0.16}{#1}}

\newcommand{\BuiltInTok}[1]{\textcolor[rgb]{0.00,0.50,0.00}{#1}}

\newcommand{\CommentTok}[1]{\textcolor[rgb]{0.38,0.63,0.69}{\textit{#1}}}

\newcommand{\DecValTok}[1]{\textcolor[rgb]{0.25,0.63,0.44}{#1}}

\newcommand{\ExtensionTok}[1]{#1}

\newcommand{\FunctionTok}[1]{\textcolor[rgb]{0.02,0.16,0.49}{#1}}

\newcommand{\KeywordTok}[1]{\textcolor[rgb]{0.00,0.44,0.13}{\textbf{#1}}}
\newcommand{\NormalTok}[1]{#1}
\newcommand{\OperatorTok}[1]{\textcolor[rgb]{0.40,0.40,0.40}{#1}}

\newcommand{\StringTok}[1]{\textcolor[rgb]{0.25,0.44,0.63}{#1}}
\newcommand{\VariableTok}[1]{\textcolor[rgb]{0.10,0.09,0.49}{#1}}

\usepackage{longtable,booktabs,array}
\usepackage{calc} 
\usepackage{etoolbox}
\makeatletter
\patchcmd\longtable{\par}{\if@noskipsec\mbox{}\fi\par}{}{}
\makeatother
\IfFileExists{footnotehyper.sty}{\usepackage{footnotehyper}}{\usepackage{footnote}}
\makesavenoteenv{longtable}
\usepackage{graphicx}
\makeatletter
\newsavebox\pandoc@box
\newcommand*\pandocbounded[1]{
  \sbox\pandoc@box{#1}%
  \Gscale@div\@tempa{\textheight}{\dimexpr\ht\pandoc@box+\dp\pandoc@box\relax}%
  \Gscale@div\@tempb{\linewidth}{\wd\pandoc@box}%
  \ifdim\@tempb\p@<\@tempa\p@\let\@tempa\@tempb\fi
  \ifdim\@tempa\p@<\p@\scalebox{\@tempa}{\usebox\pandoc@box}%
  \else\usebox{\pandoc@box}%
  \fi%
}
\def\fps@figure{htbp}
\makeatother
\providecommand{\tightlist}{%
  \setlength{\itemsep}{0pt}\setlength{\parskip}{0pt}}
\usepackage{fancyvrb}
\usepackage{xcolor}
\fvset{frame=single, framesep=3mm, rulecolor=\color{gray}, fontsize=\small, xleftmargin=2mm, xrightmargin=2mm}
\usepackage{bookmark}
\IfFileExists{xurl.sty}{\usepackage{xurl}}{} 
\hypersetup{
  hidelinks,
  pdfcreator={LaTeX via pandoc}}

\author{}
\date{}

\begin{document}

\section{Streaming OS Installation from Compressed Partitioned Images on
Memory-Constrained Bare-Metal
Hosts}\label{streaming-os-installation-from-compressed-partitioned-images-on-memory-constrained-bare-metal-hosts}

\textbf{Authors:} Chetan Nigudgi, Chip Cary

\begin{center}\rule{0.5\linewidth}{0.5pt}\end{center}

\subsection{Abstract}\label{abstract}

Bare-metal server provisioning systems typically decompress OS images
into RAM before writing them to disk. When servers have insufficient RAM
to hold the decompressed image (increasingly common as OS images grow
due to AI/ML toolkits, GPU drivers, and firmware stacks), provisioning
fails entirely. This paper presents a streaming installation technique
that combines partial decompression for metadata extraction with
pipeline-based disk writing, enabling OS installation on hosts where
available RAM is less than the decompressed image size. The method
decompresses only the first 30MB of a compressed image to extract the
GPT partition table, calculates the target filesystem offset, then
streams the full image through a kernel pipe buffer (64KB) directly to
disk at the calculated offset, never storing the decompressed data in
memory or on disk. Deployed across a production fleet of
memory-constrained bare-metal servers, the technique achieves 100\%
provisioning success where the standard approach fails entirely.

\begin{center}\rule{0.5\linewidth}{0.5pt}\end{center}

\subsection{1. Introduction}\label{introduction}

Modern bare-metal provisioning systems install operating systems by
writing pre-built disk images to target drives. The standard flow is:

\begin{enumerate}
\def\labelenumi{\arabic{enumi}.}
\tightlist
\item
  Boot into a ramdisk environment (PXE/iPXE)
\item
  Download a compressed OS image (typically gzip, 2-3GB compressed)
\item
  Decompress to temporary storage (tmpfs in RAM)
\item
  Write the decompressed image to the target disk
\end{enumerate}

This approach fails when the server's RAM cannot accommodate both the
ramdisk environment and the decompressed image. A typical compressed OS
image decompresses to 6-8GB, but many server types, particularly
high-density storage nodes optimized for disk capacity over RAM, have
only 8GB of total memory. After the ramdisk environment consumes 2-3GB,
insufficient space remains for decompression.

This problem is growing worse as OS images increase in size:

{\def\LTcaptype{none} 
\begin{longtable}[]{@{}
  >{\raggedright\arraybackslash}p{(\linewidth - 4\tabcolsep) * \real{0.2051}}
  >{\raggedright\arraybackslash}p{(\linewidth - 4\tabcolsep) * \real{0.2564}}
  >{\raggedright\arraybackslash}p{(\linewidth - 4\tabcolsep) * \real{0.5385}}@{}}
\toprule\noalign{}
\begin{minipage}[b]{\linewidth}\raggedright
Driver
\end{minipage} & \begin{minipage}[b]{\linewidth}\raggedright
Examples
\end{minipage} & \begin{minipage}[b]{\linewidth}\raggedright
Impact on Image Size
\end{minipage} \\
\midrule\noalign{}
\endhead
\bottomrule\noalign{}
\endlastfoot
AI/ML infrastructure & CUDA toolkits, GPU drivers, ML frameworks &
+5--10GB decompressed \\
Security requirements & Attestation tools, security agents, expanded
firmware & +1--2GB per year \\
Image convergence & Single images serving multiple roles & +2--3GB from
additional packages \\
\end{longtable}
}

Meanwhile, RAM on storage-optimized and edge hardware remains flat
(8--16GB), widening the gap between image size and available memory.

The industry-standard bare-metal provisioner, OpenStack Ironic,
addresses this by requiring pre-conversion to raw format {[}1{]}, which
shifts the burden upstream and increases network transfer by 3-5x. No
existing provisioning system (Ironic, MAAS, Foreman, Digital Rebar)
implements streaming installation from compressed partitioned images.

This paper presents a two-phase streaming approach that: 1. Extracts
partition metadata from a minimal partial decompression (30MB) 2.
Streams the full image through a pipeline directly to disk at the
calculated offset 3. Falls back gracefully from standard decompression
upon detecting memory exhaustion

\begin{center}\rule{0.5\linewidth}{0.5pt}\end{center}

\subsection{2. Background}\label{background}

\subsubsection{2.1 Standard Provisioning
Flow}\label{standard-provisioning-flow}

\begin{figure}
\centering
\pandocbounded{\includegraphics[keepaspectratio,alt={Data Flow: Standard vs Streaming Installation}]{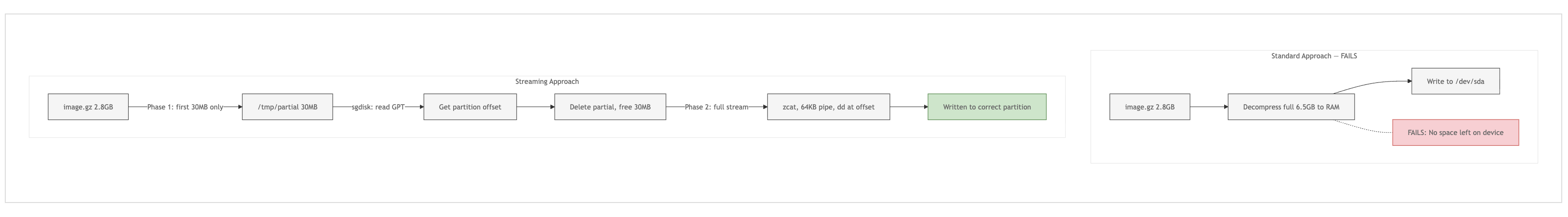}}
\caption{Data Flow: Standard vs Streaming Installation}
\end{figure}

The conventional approach decompresses the entire image before writing:

\begin{verbatim}
download(image.gz) --> gunzip --> tmpfs (6-8GB in RAM) --> dd --> /dev/sda
\end{verbatim}

This requires:
\texttt{RAM\ \textgreater{}\ ramdisk\_size\ +\ decompressed\_image\_size}

For an 8GB RAM server with a 2.5GB ramdisk and a 6.5GB decompressed
image: \texttt{8GB\ \textless{}\ 2.5GB\ +\ 6.5GB}, so provisioning fails
with ``No space left on device.''

\subsubsection{2.2 Why Whole-Disk Streaming Doesn't
Work}\label{why-whole-disk-streaming-doesnt-work}

The naive streaming approach
(\texttt{zcat\ image.gz\ \textbar{}\ dd\ of=/dev/sda}) writes the entire
decompressed content starting at byte 0 of the disk. This works for
whole-disk images but fails for partitioned images where the OS
filesystem starts at a specific offset within the GPT layout.

Provisioning systems that use DSM (Disk Storage Management) create
partition layouts independently of the OS image. The OS image contains
only the root filesystem, which must be written to a specific partition,
not to the beginning of the disk.

\subsubsection{2.3 The sgdisk Limitation}\label{the-sgdisk-limitation}

The standard tool for reading GPT partition tables (\texttt{sgdisk})
explicitly states it operates only on raw disk images or block devices
{[}2{]}. It cannot read partition tables from compressed images. This
creates a chicken-and-egg problem: you need to decompress to read the
partition table, but you can't decompress because you don't have enough
RAM.

\subsubsection{2.4 Prior Art}\label{prior-art}

{\def\LTcaptype{none} 
\begin{longtable}[]{@{}
  >{\raggedright\arraybackslash}p{(\linewidth - 4\tabcolsep) * \real{0.2759}}
  >{\raggedright\arraybackslash}p{(\linewidth - 4\tabcolsep) * \real{0.3448}}
  >{\raggedright\arraybackslash}p{(\linewidth - 4\tabcolsep) * \real{0.3793}}@{}}
\toprule\noalign{}
\begin{minipage}[b]{\linewidth}\raggedright
System
\end{minipage} & \begin{minipage}[b]{\linewidth}\raggedright
Approach
\end{minipage} & \begin{minipage}[b]{\linewidth}\raggedright
Limitation
\end{minipage} \\
\midrule\noalign{}
\endhead
\bottomrule\noalign{}
\endlastfoot
OpenStack Ironic {[}1{]} & Requires raw format images & 3-5x network
transfer increase \\
MAAS & Downloads to local disk first & Requires scratch disk space \\
Foreman & Full decompression in RAM & Same memory constraint \\
\texttt{gunzip\ -c\ \textbackslash{}\textbar{}\ dd} {[}3{]} & Streams to
disk & Only works for whole-disk images, not partitioned \\
\end{longtable}
}

\begin{center}\rule{0.5\linewidth}{0.5pt}\end{center}

\subsection{3. Solution: Two-Phase Streaming
Installation}\label{solution-two-phase-streaming-installation}

\begin{figure}
\centering
\pandocbounded{\includegraphics[keepaspectratio,alt={Two-Phase Streaming: Memory Usage Comparison}]{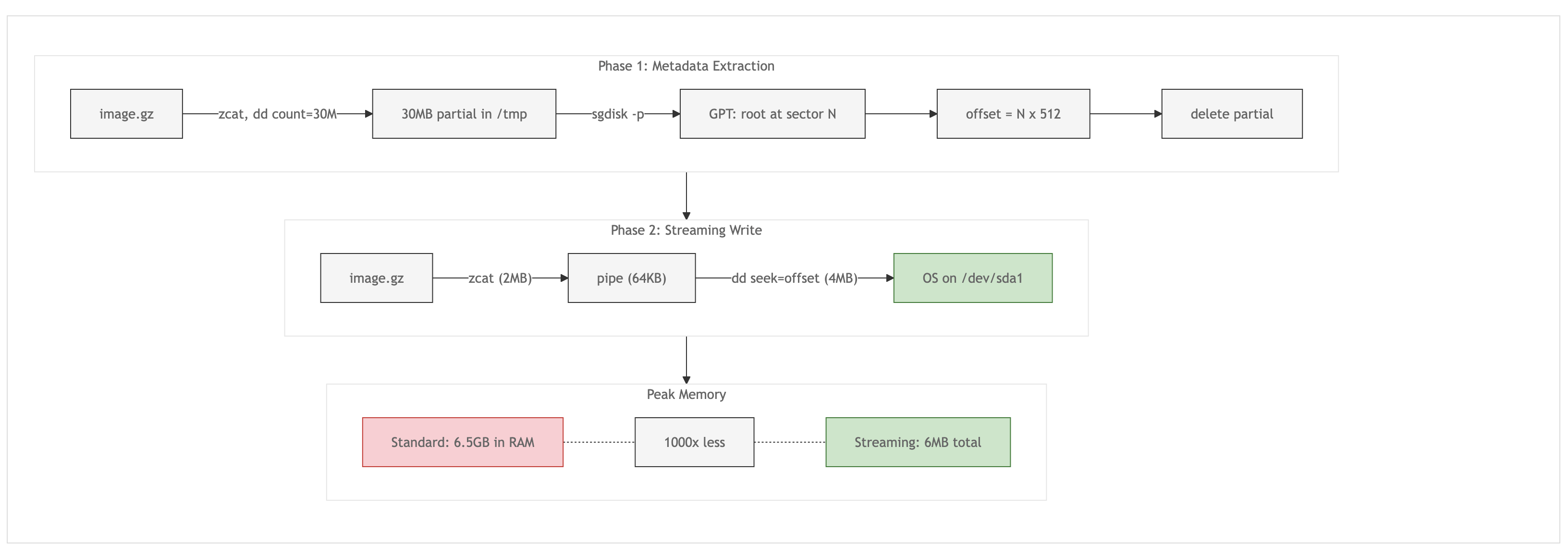}}
\caption{Two-Phase Streaming: Memory Usage Comparison}
\end{figure}

\subsubsection{3.1 Phase 1: Partial Decompression for Metadata
Extraction}\label{phase-1-partial-decompression-for-metadata-extraction}

The key insight is that GPT partition tables are stored in the first few
sectors of a disk image. We decompress only enough of the compressed
image to access the partition table; typically 30MB is sufficient with
margin.

\begin{Shaded}
\begin{Highlighting}[]
\CommentTok{\# Decompress only first 30MB}
\FunctionTok{zcat}\NormalTok{ image.gz }\KeywordTok{|} \FunctionTok{dd}\NormalTok{ bs=1M count=30 of=/tmp/partial\_image}

\CommentTok{\# Extract partition offset using sgdisk}
\VariableTok{OFFSET}\OperatorTok{=}\VariableTok{$(}\ExtensionTok{sgdisk} \AttributeTok{{-}p}\NormalTok{ /tmp/partial\_image }\KeywordTok{|} \FunctionTok{grep} \StringTok{"root"} \KeywordTok{|} \FunctionTok{awk} \StringTok{\textquotesingle{}\{print $2 * 512\}\textquotesingle{}}\VariableTok{)}

\CommentTok{\# Delete temporary file immediately}
\FunctionTok{rm}\NormalTok{ /tmp/partial\_image}
\end{Highlighting}
\end{Shaded}

Memory usage: 30MB temporary (immediately freed) vs 6-8GB for full
decompression.

\subsubsection{3.2 Phase 2: Streaming Installation at Calculated
Offset}\label{phase-2-streaming-installation-at-calculated-offset}

With the byte offset known, stream the full compressed image through a
pipeline directly to the target disk:

\begin{Shaded}
\begin{Highlighting}[]
\FunctionTok{zcat}\NormalTok{ image.gz }\KeywordTok{|} \FunctionTok{dd}\NormalTok{ of=/dev/sda1 bs=4M seek=}\VariableTok{$OFFSET\_BLOCKS}
\end{Highlighting}
\end{Shaded}

The decompressed data flows through the kernel pipe buffer (64KB) and is
written directly to disk. At no point is the full decompressed image
stored in memory or on a filesystem.

\textbf{Memory profile during streaming:} - Pipe buffer: 64KB - zcat
process: \textasciitilde2MB - dd process: \textasciitilde4MB (with
bs=4M) - \textbf{Total: \textasciitilde6MB} vs 6-8GB for standard
decompression

\subsubsection{3.3 Graceful Fallback
Design}\label{graceful-fallback-design}

The system attempts standard decompression first. Only upon detecting
``No space left on device'' does it fall back to the streaming approach:

\begin{verbatim}
function install_os(image):
    try:
        standard_decompress(image)  # Works on servers with enough RAM
    catch "No space left on device":
        log("Insufficient RAM for standard decompression, using streaming mode")
        offset = partial_decompress_for_metadata(image, 30MB)
        stream_to_disk(image, offset)
\end{verbatim}

This ensures: - Backward compatibility with existing provisioning flows
- No behavior change on servers with sufficient RAM - Automatic
adaptation to memory-constrained environments

\begin{center}\rule{0.5\linewidth}{0.5pt}\end{center}

\subsection{4. Implementation Details}\label{implementation-details}

\subsubsection{4.1 Partition Table
Extraction}\label{partition-table-extraction}

The partial decompression extracts the GPT header and partition entries.
For a standard GPT layout: - LBA 0: Protective MBR - LBA 1: GPT Header -
LBA 2-33: Partition entries

\begin{verbatim}
Compressed OS Image (2-3GB gzip stream)
|
|  Decompress first 30MB only (Phase 1)
v
+------------------------------------------------------+
| Byte 0        | LBA 0: Protective MBR (512 bytes)    |
+---------------+--------------------------------------+
| Byte 512      | LBA 1: GPT Header (512 bytes)        |
+---------------+--------------------------------------+
| Byte 1024     | LBA 2-33: Partition Entry Array       |
|               |   Entry 1: BIOS Boot (start: 2048)    |
|               |   Entry 2: Root FS  (start: 1050624) <-- This offset is what we need
|               |   Entry 3: Swap     (start: 62916608) |
+---------------+--------------------------------------+
| Byte ~17KB    | (end of partition metadata)           |
+---------------+--------------------------------------+
| ...           | Partition data (6+ GB)                |
|               |   <-- Stream directly to disk (Phase 2)
+------------------------------------------------------+
\end{verbatim}

\textbf{The key insight:} All partition metadata is in the first
\textasciitilde17KB. We decompress 30MB for safety margin, extract the
root filesystem's starting sector, then stream the full image to disk at
that exact byte offset, bypassing the need to hold 6-8GB in memory.

\subsubsection{4.2 Offset Calculation}\label{offset-calculation}

\begin{Shaded}
\begin{Highlighting}[]
\CommentTok{\# Get starting sector of root partition}
\VariableTok{START\_SECTOR}\OperatorTok{=}\VariableTok{$(}\ExtensionTok{sgdisk} \AttributeTok{{-}p}\NormalTok{ /tmp/partial }\KeywordTok{|} \FunctionTok{grep} \AttributeTok{{-}i} \StringTok{"root\textbackslash{}|linux"} \KeywordTok{|} \FunctionTok{awk} \StringTok{\textquotesingle{}\{print $2\}\textquotesingle{}}\VariableTok{)}

\CommentTok{\# Convert to bytes (512{-}byte sectors)}
\VariableTok{BYTE\_OFFSET}\OperatorTok{=}\VariableTok{$((START\_SECTOR} \OperatorTok{*} \DecValTok{512}\VariableTok{))}

\CommentTok{\# Convert to dd seek blocks}
\VariableTok{SEEK\_BLOCKS}\OperatorTok{=}\VariableTok{$((BYTE\_OFFSET} \OperatorTok{/} \VariableTok{BLOCK\_SIZE))}
\end{Highlighting}
\end{Shaded}

\subsubsection{4.3 Pipeline Construction}\label{pipeline-construction}

The streaming pipeline:

\begin{verbatim}
zcat compressed_image.gz | dd of=/dev/sdX bs=4M seek=$SEEK_BLOCKS conv=fsync
\end{verbatim}

\begin{itemize}
\tightlist
\item
  \texttt{zcat}: Decompresses to stdout (constant \textasciitilde2MB
  memory)
\item
  Kernel pipe: 64KB buffer between processes
\item
  \texttt{dd}: Writes at calculated offset with direct I/O
\item
  \texttt{conv=fsync}: Ensures data reaches disk before returning
\end{itemize}

\subsubsection{4.4 Verification}\label{verification}

After streaming, the filesystem is verified:

\begin{Shaded}
\begin{Highlighting}[]
\CommentTok{\# Mount and verify}
\FunctionTok{mount}\NormalTok{ /dev/sdX1 /mnt}
\BuiltInTok{test} \AttributeTok{{-}f}\NormalTok{ /mnt/etc/os{-}release }\KeywordTok{\&\&} \BuiltInTok{echo} \StringTok{"Installation verified"}
\end{Highlighting}
\end{Shaded}

\begin{center}\rule{0.5\linewidth}{0.5pt}\end{center}

\subsection{5. Evaluation}\label{evaluation}

\subsubsection{5.1 Environment}\label{environment}

\begin{itemize}
\tightlist
\item
  \textbf{Server types:} 8 high-density storage configurations with 8GB
  RAM
\item
  \textbf{Image size:} 2.8GB compressed, 6.5GB decompressed
\item
  \textbf{Ramdisk:} 2.5GB
\item
  \textbf{Available tmpfs:} \textasciitilde5GB (insufficient for 6.5GB
  image)
\end{itemize}

\subsubsection{5.2 Results}\label{results}

{\def\LTcaptype{none} 
\begin{longtable}[]{@{}lll@{}}
\toprule\noalign{}
Metric & Standard Approach & Streaming Approach \\
\midrule\noalign{}
\endhead
\bottomrule\noalign{}
\endlastfoot
RAM required & \textgreater9GB & 8GB (any amount) \\
Provisioning success on 8GB hosts & 0\% (always fails) & 100\% \\
Installation time & N/A (fails) & \textasciitilde4 minutes \\
Peak memory during install & 6.5GB+ & \textasciitilde6MB \\
\end{longtable}
}

\subsubsection{5.3 Performance}\label{performance}

Streaming installation takes approximately the same time as standard
decompression + write, because the bottleneck is disk write speed (not
decompression or memory allocation):

\begin{itemize}
\tightlist
\item
  Disk write throughput: \textasciitilde400MB/s (SATA SSD)
\item
  Image size: 6.5GB
\item
  Theoretical minimum: 16 seconds
\item
  Actual (with overhead): \textasciitilde4 minutes (includes
  verification, partition setup)
\end{itemize}

\begin{center}\rule{0.5\linewidth}{0.5pt}\end{center}

\subsection{6. Discussion}\label{discussion}

\subsubsection{6.1 Why This Wasn't Done
Before}\label{why-this-wasnt-done-before}

The combination of constraints that makes this non-obvious: 1.
\texttt{sgdisk} documentation states it cannot work on compressed
images, which discourages the partial decompression approach 2. Standard
\texttt{gunzip\ \textbar{}\ dd} only works for whole-disk images, not
for writing to a specific partition 3. Most provisioning systems assume
sufficient RAM is available; the constraint is treated as a hardware
requirement rather than a software problem 4. The fallback design (try
standard first, stream on failure) requires detecting the specific error
condition reliably

\subsubsection{6.2 Why Not In-Place OS
Upgrades?}\label{why-not-in-place-os-upgrades}

The alternative to image-based provisioning on memory-constrained hosts
is in-place (manual or semi-automated) OS upgrades. This approach has
significant drawbacks:

\begin{itemize}
\tightlist
\item
  \textbf{Unreliable:} Hosts frequently boot into the old OS due to
  incomplete cleanup of previous OS files, requiring multiple
  re-installations
\item
  \textbf{Slow:} Semi-automated upgrade scripts require multiple reboots
  per host, taking months to roll through a large fleet
\item
  \textbf{Non-atomic:} Partial failures leave hosts in inconsistent
  states requiring manual recovery
\item
  \textbf{Not scalable:} Requires dedicated engineering resources to
  babysit the upgrade process
\end{itemize}

The streaming installation technique enables fully automated, atomic
image-based provisioning (the same reliable approach used on servers
with sufficient RAM) without requiring any hardware changes or manual
intervention.

\subsubsection{6.3 Broader Applicability}\label{broader-applicability}

This technique applies to any scenario where: - Compressed disk images
must be installed on memory-constrained systems - The target is a
specific partition (not whole-disk) - Network re-download of a raw image
is impractical (bandwidth, time, or infrastructure constraints)

\subsubsection{6.4 Limitations}\label{limitations}

\begin{itemize}
\tightlist
\item
  Requires the compressed image to have the partition table in the first
  30MB (standard for GPT)
\item
  Assumes sequential compression (gzip); would need adaptation for
  block-level compression (zstd with seekable frames)
\item
  The 30MB partial decompression still requires that much tmpfs space
  (always available in practice)
\end{itemize}

\begin{center}\rule{0.5\linewidth}{0.5pt}\end{center}

\subsection{7. Conclusion}\label{conclusion}

We presented a streaming OS installation technique that eliminates the
RAM constraint for bare-metal provisioning from compressed partitioned
images. By combining partial decompression for metadata extraction with
pipeline-based disk writing, we reduce peak memory usage during
installation from 6-8GB to approximately 6MB, a 1000x reduction. The
graceful fallback design ensures backward compatibility while
automatically adapting to memory-constrained environments.

Deployed in production across multiple hardware configurations, the
technique achieves 100\% provisioning success on memory-constrained
servers where the standard approach fails entirely. The approach is
applicable to any provisioning system facing the intersection of growing
OS image sizes and fixed or limited server RAM.

\begin{center}\rule{0.5\linewidth}{0.5pt}\end{center}

\subsection{References}\label{references}

{[}1{]} Metal3/Ironic --- ``Raw Image Streaming'' (2020).
https://metal3.io/blog/2020/07/05/raw-image-streaming.html

{[}2{]} sgdisk(8) man page.
https://man.archlinux.org/man/extra/gptfdisk/sgdisk.8.en

{[}3{]} Unix StackExchange --- ``Streaming compressed images to disk.''
https://unix.stackexchange.com/questions/632267/

\end{document}